\documentclass[aps,pre,twocolumn,groupedaddress,superscriptaddress,showpacs]{revtex4-1}
\usepackage{tikz}
\usepackage{amsmath}
\usepackage{graphicx}
\usepackage{float}
\usepackage{color}
\usepackage{comment}
\usepackage{tikz}
\usepackage{gensymb}
\usepackage{siunitx}
\usepackage{float}
\usepackage{scalerel,amssymb}

\usepackage{pifont}

\begin{document}
	
	\title{ Wetting of a Solid Surface by Active Matter} 

	\author{P. D. Neta}
    \email{pedrodidier@hotmail.com}
    \affiliation{Departamento de F\'{\i}sica, Faculdade de Ci\^{e}ncias, Universidade de Lisboa, 
    1749-016 Lisboa, Portugal}
    \affiliation{Centro de F\'{i}sica Te\'{o}rica e Computacional, Faculdade de Ci\^{e}ncias, Universidade de Lisboa, 1749-016 Lisboa, Portugal}

    \author{M. Tasinkevych}
    \email{mtasinkevych@fc.ul.pt}
    \affiliation{Departamento de F\'{\i}sica, Faculdade de Ci\^{e}ncias, Universidade de Lisboa, 
    1749-016 Lisboa, Portugal}
    \affiliation{Centro de F\'{i}sica Te\'{o}rica e Computacional, Faculdade de Ci\^{e}ncias, Universidade de Lisboa, 1749-016 Lisboa, Portugal}
    
    \author{M. M. Telo da Gama}
    \email{mmgama@fc.ul.pt}
    \affiliation{Departamento de F\'{\i}sica, Faculdade de Ci\^{e}ncias, Universidade de Lisboa, 
    1749-016 Lisboa, Portugal}
    \affiliation{Centro de F\'{i}sica Te\'{o}rica e Computacional, Faculdade de Ci\^{e}ncias, Universidade de Lisboa, 1749-016 Lisboa, Portugal}
    
    \author{C. S. Dias}
    \email{csdias@fc.ul.pt}
    \affiliation{Departamento de F\'{\i}sica, Faculdade de Ci\^{e}ncias, Universidade de Lisboa, 
    1749-016 Lisboa, Portugal}
    \affiliation{Centro de F\'{i}sica Te\'{o}rica e Computacional, Faculdade de Ci\^{e}ncias, Universidade de Lisboa, 1749-016 Lisboa, Portugal}

\begin{abstract}

 A lattice model is used to study repulsive active particles at a planar surface. A rejection-free Kinetic Monte Carlo method is employed to characterize the wetting behaviour. The model predicts a mobility induced phase separation of active particles, and the bulk coexistence of dense liquid-like and dilute vapour-like steady states is determined. An ``ensemble'', with a varying number of particles, analogous to a grand canonical ensemble in equilibrium, is introduced. The formation and growth of the liquid film between the solid surface and the vapour phase is investigated. For all of the activities considered, the thickness of the adsorbed film exhibits a diverging behaviour as the system is brought towards coexistence from the vapour side, suggesting a complete wetting scenario along the full coexistence curve.


\end{abstract}

\maketitle
 
%--------------------------- INTRO--------------------------------	
\section{Introduction}

In the last decade, significant progress has been made in developing synthetic self-propelled particles \cite{Bechinger2016,Patra2013}, which are envisioned as autonomous micromotors in  several new applications, such as cargo transport at the microscale \cite{Baraban2012,Patra2013,Xu2020}, the assembly of structures via autonomous local deposition of materials \cite{Sanchez2015} or pollution remediation systems \cite{Soler2013,Soler2014}. Active Janus colloids form a large class of active particles employing self-phoresis \cite{anderson89,golestanian05,Howse2007,golestanian07,jiang10,Moran11,Moran17,Illien17}, i.e., the generation of a tangential surface flow powered by, e.g., light-induced local temperature gradients \cite{jiang10}, chemical reaction-generated composition gradients in solutions \cite{Howse2007,golestanian07,popescu10} or electric potentials  \cite{Moran11,ebbens14,brown14,Brown17,Moran17}. Janus colloidal spheres can also move through the generation of composition gradients in near critical mixtures, illuminated by light \cite{Bechinger2012}.

 From an academic viewpoint, active colloids are non-equilibrium systems that violate detailed balance, and exhibit novel types of self-organization, such as “living crystals” that are mobile, break apart and reform again \cite{Palacci2013}. Computer simulations of active Brownian particles (ABP), with repulsive interactions, revealed the existence of a phase separation between non-equilibrium steady states with different number densities of the particles \cite{Fily2012,Redner2013,Levis2014}. This phase separation was predicted by Tailleur and Cates for run-and-tumble particles with a motility that decreases sufficiently fast with increasing local density \cite{Tailleu2008}. The transition is driven by the activity of the repulsive particles and is known as motility-induced phase separation (MIPS). As the free self-propulsion speed decreases, the density difference between the coexisting phases decreases and it vanishes at a critical point \cite{Siebert2018}. Evidence for MIPS was also reported in experiments \cite{Buttinoni2013} on two-dimensional suspensions of self-propelled carbon-coated Janus particles in a near-critical water-lutidine mixture. The underlying  physical mechanism is related to the fact that active particles slow down, due to the collisions with other particles, in crowded regions, and that they tend to accumulate where they move slower \cite{Schnitzer1993}. This positive feedback leads to the accumulation of active particles in regions with higher density and ultimately drives phase separation \cite{Cates2015}. 

Continuum equations for the conserved local density of active particles and the local polarization order parameter were derived by coarse-graining the microscopic dynamics given by the Smoluchowski equation for the many-body probability distribution function (PDF) of the particles positions and orientations \cite{Fily2012,Bialk2013,Cates2013,Stenhammar2013,Bialk2013,Speck2014,Speck2015,Cates2015}. Alternatively, a continuum description may be obtained from phenomenological considerations, based on symmetry arguments and conservation laws, where the governing equations for a set of properly defined order parameters are written as an expansion in powers of the order parameters and their gradients \cite{Wittkowski2014,Cates2015,Solon2018}. Detailed analysis of the coarse-grained continuum models revealed that on large length scales (larger than the persistence length of the directed motion of the active particles \cite{Speck2015}) the dynamics of ABPs can be mapped onto the equilibrium dynamics of a system of passive particles with attractive interactions \cite{Speck2014,Speck2015,Cates2015}, with a corresponding effective free energy of the classical Ginzburg-Landau form \cite{Speck2015}. A theoretical method to calculate the activity-induced effective interactions, by integration of the orientational degrees of freedom of ABPs, was reported in ref.~\cite{Farage2015}. A Langevin equation for the particles positions with colored noise was found, and a perturbative expansion in powers of the noise correlation time renders an approximate Fokker-Planck equation for the PDF of the positional degrees of freedom. From this equation an effective activity-dependent particle potential can be inferred. Later on, the range of applicability of this mapping was tested in ref.~\cite{Rein2016} by comparing the results of simulations of active Brownian particles and a system of passive particles with the effective interactions as predicted in ref.~\cite{Farage2015}. The conclusion was ``that beyond linear response, active particles exhibit genuine non-equilibrium properties that cannot be captured by the effective pair interaction alone'' \cite{Rein2016}. 

The coexistence between steady states of ABPs can also be understood from the continuity equation governing the dynamics of the local particle density, without the need for an effective free energy. For instance, in a spatially inhomogeneous steady state with planar geometry the particle flux must vanish. This can be used to formally define scalar analogues of a local chemical potential and a local pressure which must be uniform across the coexisting phases at a planar interface \cite{Paliwal2018}. Further insights into the structure of these generalized chemical potential and pressure were gained from the power functional theory developed by Schmidt and coworkers \cite{Schmidt2013,Krinninger2019,Hermann2019b}. This microscopic theory dictates that the generalized chemical potential (and pressure) is the sum of an adiabatic term, which depends only on the density profile and can be obtained from the excess free energy functional of the corresponding adiabatic system, and a genuine nonequilibrium superadiabatic term, which depends both on the density and current profiles \cite{Hermann2019a,Hermann2019b}.  Despite some progress, understanding the collective dynamics of active particles is one of the grand challenges in the field of active matter. The problem is particularly difficult not only due to its intrinsic non-equilibrium character but also to the complex nature of the effective interparticle interactions, where a subtle interplay of hydrodynamic and phoretic effects \cite{Liebchen2019a,Popescu2019,Liebchen2019b,Kanso2019} requires innovative theoretical approaches and numerical methods. 
  
Significant efforts have been directed towards understanding the surface tension of an interface between two coexisting steady states of ABPs. In the case of equilibrium interfaces, the  surface tension can be defined following several alternative routes, mechanical or thermodynamic,  all of which lead to the same result \cite{Paliwal2017}. However, this is not the case for the effective surface tension of non-equilibrium interfaces in systems that have undergone MIPS. Simulation results revealed that the surface tension obtained via integration of the anisotropy of the stress tensor is negative \cite{Bialke2015, Marconi2015,Patch2018,Solon2018}. This behavior has been rationalized in terms of the surface polarization and the associated contributions to the active stress tensor at the interface \cite{Speck2020}. By contrast, an effective thermodynamic route, based on insights from the power functional theory, predicts positive surface tensions \cite{Hermann2019a}. Finally, a generalized thermodynamic framework developed in ref.~\cite{Solon2018} supports both signs for the surface tension. 

Recently, adsorption of active particles onto a solid surface was reported in refs.~\cite{Sepulveda2017,Sepulveda2018}, where it was demonstrated that the thickness of the adsorbed liquid layer of active particles increases with decreasing turning rate of the particles. The turning rate was set either by the tumbling rate of run-and-tumble lattice models \cite{Sepulveda2017}, or by the rotational diffusion coefficient of off-lattice ABPs models \cite{Sepulveda2018}. However, these studies did not address MIPS, rendering the interpretation of the results as wetting transitions questionable.

Here we report a simulation study of the interface between a planar solid surface and a low density phase of ABPs undergoing MIPS. For brevity, we refer in the following to the low density phase as the "vapour" phase of the ABPs, and to the high density phase as the "liquid" phase. Our objective is to study wetting of the solid-vapor interface by the liquid phase. More specifically we want to understand whether the solid-vapor interface exhibits a wetting transition in a classic interfacial thermodynamic sense \cite{Gennes1985,Sullivan1986, dietrich1988,Forgacs1991}, i.e. whether the solid-vapor interface can be separated by a macroscopically thick liquid layer. At equilibrium, the wetting transition is associated with a singularity of the surface excess free energy, at the liquid-vapor bulk coexistence curve \cite{Sullivan1986,dietrich1988}. Below the wetting temperature, the solid-vapor interface is formed by a microscopically thin liquid film, or equivalently the liquid forms a sessile droplet on the solid surface with a non zero contact angle \cite{Sullivan1986, dietrich1988}. With increasing temperature along the liquid-vapour coexistence, the thickness of the liquid film adsorbed on the substrate grows (the contact angle of the sessile droplet decreases) and eventually diverges (the contact angle approaches zero) at the wetting temperature \cite{Sullivan1986,dietrich1988}. The film thickness can diverge either continuously (second order wetting transition) or discontinuously (first order wetting transition) \cite{Sullivan1986, dietrich1988}. A heuristic argument for the existence of the wetting transition at the solid-vapour interface follows from an elegant analysis due to Cahn \cite{Cahn1977}, based on the fact that close to the critical point the liquid-vapor surface tension $\sigma_{lv}$ as a function of the reduced temperature vanishes faster than the vanishing difference (as a function of the reduced temperature) of the solid-vapor $\sigma_{sv}$ and the solid-liquid   $\sigma_{sl}$ surface tensions. From this it follows that there exists a temperature $T_w$ where $\sigma_{lv}=\sigma_{sv}-\sigma_{sl}$ holds. The last equation is equivalent to the vanishing of the contact angle of a sessile liquid droplet \cite{Sullivan1986, dietrich1988}, or complete wetting.

In what follows, we apply a rejection-free Kinetic Monte Carlo (KMC) method to a lattice model of active particles. In the first step we calculate the bulk coexistence curve of MIPS. Then we introduce a quasi-grand canonical realization of the KMC method in order to investigate the wetting behaviour of the solid-vapour interface. We calculate the thickness of the wetting film upon approaching the phase transition from the vapour side at a constant value of the activity. This mimics an isothermal approach to a phase boundary in equilibrium systems. The film thickness as a function of the distance to the bulk coexistence curve does not show marked differences at different activities. In all cases we observed signatures of a diverging film thickness close to coexistence. A more detailed analysis of the simulation results suggests that this divergence could be of a double logarithm type, as in equilibrium wetting with short range attractive forces, for all the activities $\epsilon$ where MIPS is observed. Although the precise form of the divergence is difficult to ascertain by simulation, our results indicate that the wetting film diverges in the whole range of $\epsilon$, i.e. that the solid surface is always wet at the MIPS phase boundary. In equilibrium terms this corresponds to a wetting transition at zero temperature, as observed in simple models where the liquid phase persists down to zero temperature. 
      
%--------------------------- MODEL-------------------------------- 
\section{Model}

\subsection{Canonical}

\begin{figure}[h!]
    %\captionsetup{justification=raggedright,singlelinecheck=false}
    \centering
    \includegraphics[width=190pt, height=160pt]{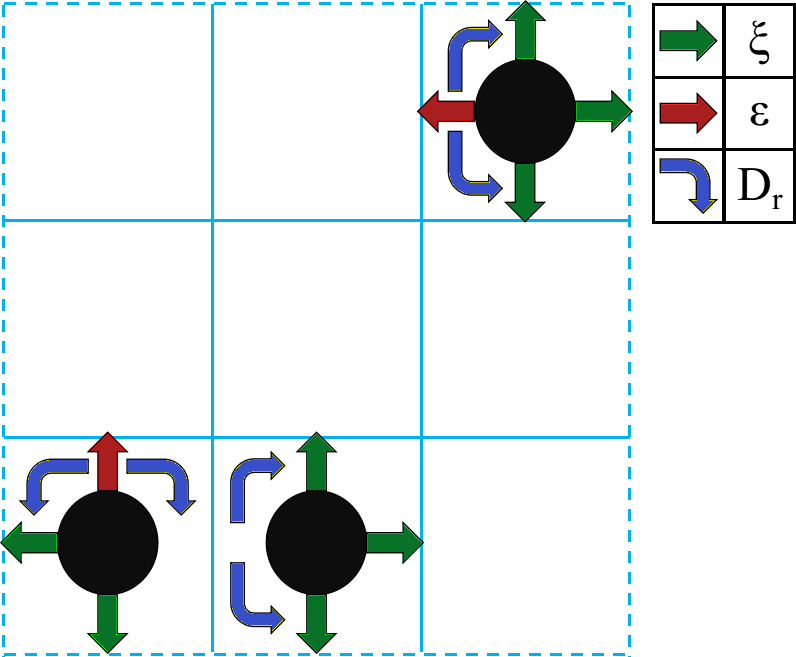}
    \caption{\small{$3 \times 3$ lattice, with a dashed boundary representing periodic boundary conditions (PBC), and 3 active particles interacting via excluded volume. At each time step, one particle can either translate to an empty nearest-neighbour lattice site or rotate. The green, red and yellow arrows represent the active propulsion and the thermal (diffusive) translation and rotation, respectively. The arrow that represents active or diffusive translation to a nearest neighbour site is removed if that site is occupied, while the arrow that depicts rotation is kept as the particle may rotate even in a cluster configuration.}}
    \label{lattice_representation}
\end{figure}

We use a two-dimensional lattice model to simulate the MIPS and the wetting behaviour of a solid surface. The MIPS phase diagram is obtained for systems with a fixed number of particles (N) while the wetting behaviour is investigated through the growth of the adsorbed liquid films for open systems, at constant activity as the coexistence is approached. These conditions resemble (but are not identical to) those of equilibrium systems in the canonical (closed system) and the grand canonical (open system) ensembles, as discussed below. 
We use simulation boxes of size $L_x \times L_y$ and consider $N$ active particles interacting through excluded volume only, setting the global density $\rho_t=\frac{N}{L_x \times L_y}$. The translational dynamics of the individual particles is governed by self-propulsion and thermal diffusion. In addition to translational diffusion, the particles also diffuse rotationally, changing the self-propelling direction by $\pm\frac{\pi}{2}$, endowing the system with a short-time memory. A typical lattice is depicted in Fig. \ref{lattice_representation} with a schematic representation of the dynamics of the active particles.

%figure---------------------------------------------- LATTICE REPRESENTATION

We implemented a rejection-free KMC method to simulate efficiently the time evolution of the system \cite{Bortz1975}. Three processes are considered, each with its specific rate: thermal diffusion ($\xi$), active motion ($\epsilon$), and rotational diffusion ($D_r$). We set both the lattice spacing and the thermal diffusion coefficient to unity and rescale all the other quantities accordingly, leaving the activity and the rotational diffusion coefficient as free parameters ($\xi = 1, \epsilon, D_r$). When $\epsilon=1$ there is no activity and the system is passive, while at high activities $\epsilon \gg 1$, the particles move persistently and almost exclusively in the self-propelling direction (in finite lattices). 

\begin{comment}

\end{comment}

In order to simulate the time evolution of the system using KMC, we start by calculating the total rate at each time step $t$, which is the sum of the rates of all the processes, for all the particles:

\begin{equation}
    R_{total}=  2ND_r + N_- + N_{+} \epsilon , 
\end{equation}
where $N_+$ and $N_-$ represent the number of self-propelling and thermal diffusion translational events, respectively. As each particle can rotate both clockwise and counterclockwise at any time, $D_r$ is multiplied by $2N$. Given the total rate, at each time step a process is chosen at random, according to its weight, and one of the particles for which the process is allowed, also chosen at random, realizes it. The lattice is updated, the process rates are recalculated and the procedure is repeated up to a stopping condition. After each iteration, the time evolution is incremented by

%After determining  $R_{total}$, a random process is chosen with probability $r_t/R_{total}$, where $r_t$ is the rate of the process $r$ at time $t$.

\begin{equation}
    \Delta t = \frac{-1}{R_{total}} \ln(\eta) \hspace{+0.2cm} , \hspace{+0.2cm} \eta \in ]0:1]  . 
    \label{eq::KMC_time_increment}
\end{equation}
where $\eta$ is a number chosen at random in the interval $]0,1]$. 

We then proceed to calculate the MIPS phase diagram reported in Section~\ref{sec.res_can}.

%%%%%%%%%%%%%%%%%%%%%%%%%%%%%%%%%%%%%%%%%%%%%%%%%%%%%%%%%%%%% second model
\subsection{\textit{Quasi} Grand Canonical}

\begin{figure}[h!]
        %\captionsetup{justification=raggedright,singlelinecheck=false}
    	\includegraphics[width=210pt, height=170pt]{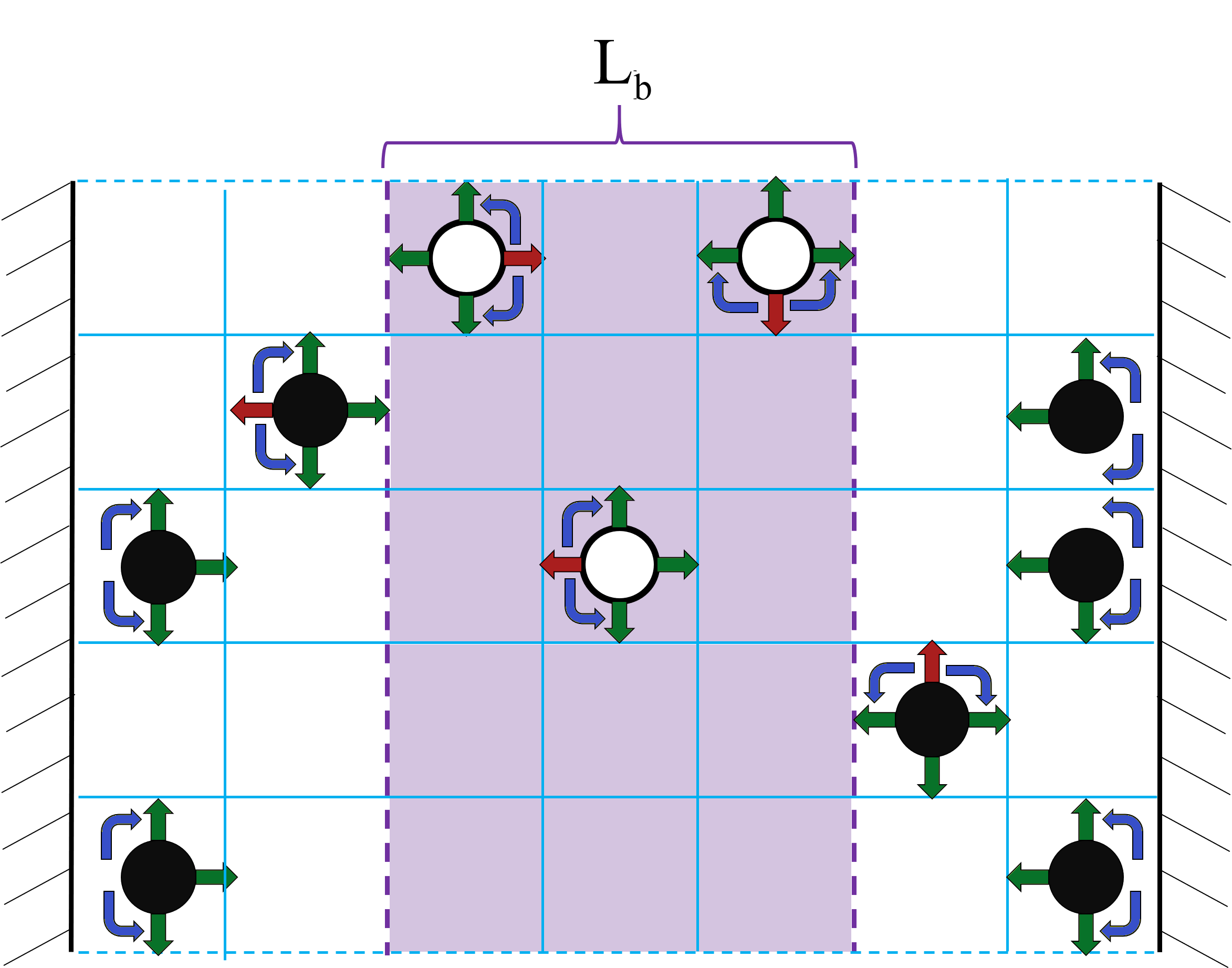}
        \caption{\small{Representation of a lattice in the \textit{Quasi} Grand Canonical ensemble. Particles in the bulk region (purple) may be inserted or removed to keep the vapour density fixed. After insertion, white particles in the bulk reservoir follow the dynamics of the model. Black particles follow the dynamics of the model. The particles will change colour if they cross the reservoir boundaries (dashed lines). The solid vertical lines represent the solid surfaces and PBC are used in the y-direction. }}
        \label{lattice_representation_inserting}
\end{figure}

In order to account for the solid surface, we use closed boundary conditions in the x-direction (see Fig. \ref{lattice_representation_inserting}) effectively simulating two solid-vapour interfaces.

At complete wetting, we expect an infinite liquid film of active particles to grow at the solid-vapour surface, under conditions of bulk coexistence. In order to simulate this behaviour we have to consider an open system, coupled to an infinite reservoir at the coexisting vapour density. At equilibrium, we avoid the simulation of the reservoir by using the grand-canonical ensemble. In active systems a finite but very large reservoir has been simulated explicitly  \cite{ni2015} but this is a very expensive way to simulate open systems. The difficulty in implementing a grand-canonical simulation for active systems stems from the fact that these systems are far from equilibrium and (i) the equivalence of different ensembles is not guaranteed, and (ii) in general, there are difficulties in defining the chemical potential, as mentioned briefly in the introduction.   
 
In order to keep the bulk density fixed as the liquid film grows at the surfaces, we insert and remove active particles in the bulk region far away from the surfaces, ensuring that the bulk vapour density is fixed throughout the simulation. In our system, the bulk 
is located in the center of the simulation box, and has a length $L_b=200$ (in most  simulations), which is sufficient to consider that the two solid-vapour interfaces are uncorrelated and that the insertion and removal of particles does not affect the dynamics of the film growth. The implementation of this \textit{Quasi} Grand Canonical ensemble allows the simulation of growing films at surfaces, at constant bulk density, without the need for expensive simulations of the particle reservoir \cite{ni2015}.

To analyse the film growth at the surface and its interface with the bulk vapour, we measure the interfacial height ($h$) using the burning method \cite{Herrmann1984}. We consider the interface to be the line that separates the set of particles that are directly in contact or connected through nearest neighbors to the surface, from the remaining particles.

%figure----------------------------------------------

%--------------------------- RESULTS-------------------------------- 

\section{Results}
\subsection{Canonical}\label{sec.res_can}

%Fig.---------------------------------------------- Density histogram comparison
\begin{figure}[h!]
  \centering
  \begin{tabular}{@{}c@{}}
    \vspace{-.0cm}
    \includegraphics[height=0.999\linewidth,width=80pt,angle=270]{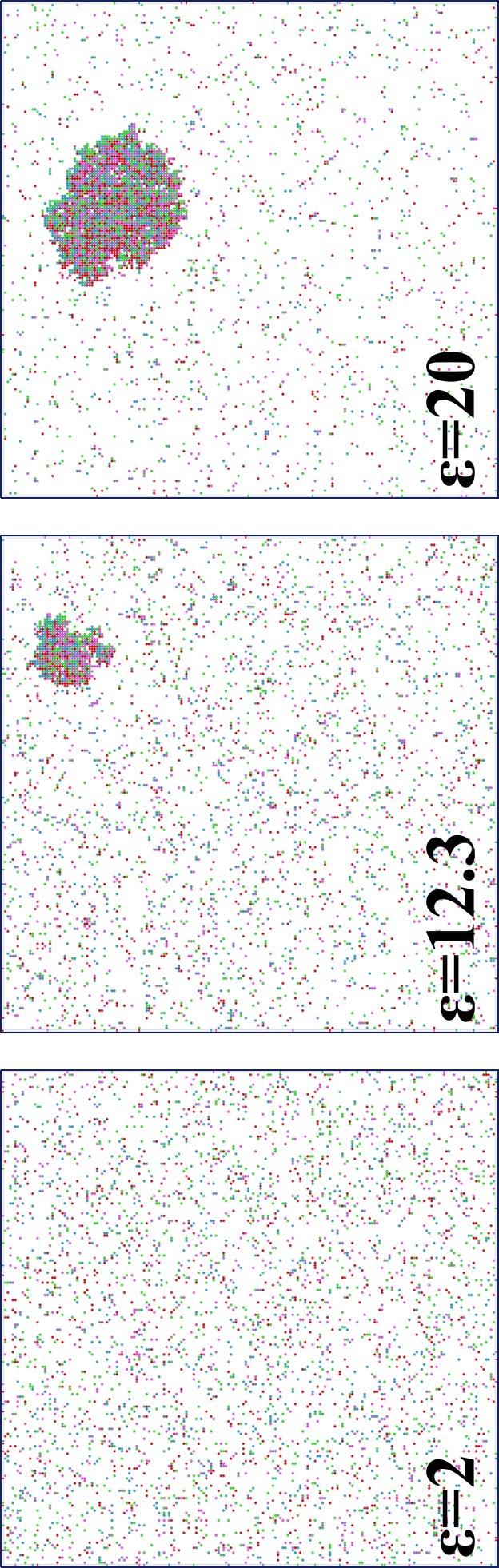} \\[\abovecaptionskip]
  \end{tabular}
  \begin{tabular}{@{}c@{}}
    \vspace{-.0cm}
    \includegraphics[width=0.9\columnwidth]{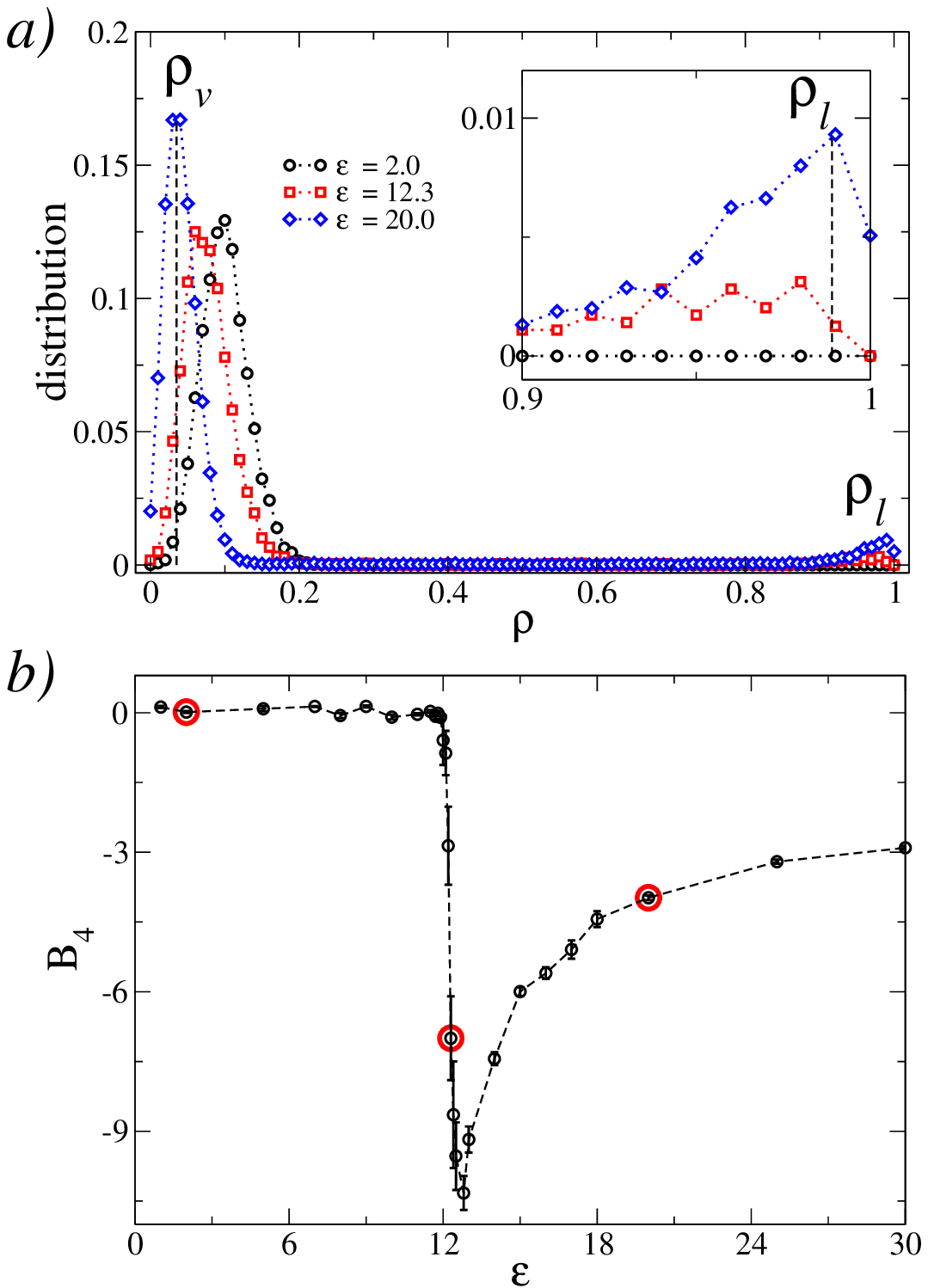} \\[\abovecaptionskip]
  \end{tabular}
       \caption{\small{Results for the bulk simulation at different activities, in a box with $L = 200$, and fixed $D_r=0.1$ and $\rho_t = 0.1$. \textit{\textbf{Top panel:}} Snapshots at various activities: $\epsilon= 2$, $\epsilon= 12.3$ and $\epsilon= 20$; Clustering is observed at $\epsilon \geq$ 12.3, suggesting MIPS. Particles oriented to the right, up, left and down are coloured in red, pink, green and blue, respectively. 
        \textbf{\textit{a)}} Local density distributions $\rho$. At $\epsilon=2$, the single peak is at $0.1$ (black dotted line) as in passive systems. When the activity increases to $\epsilon=12.3$, a second high density peak appears, signalling the presence of a high density cluster. As the activity increases the cluster size increases and the distinction between the high and low density phases becomes clearer. The curves are an average over 50 samples. \textbf{\textit{b)}} Fourth order Binder cumulant ($B_4$) as a function of the activity ($\epsilon$). When the activity is lower than $ \epsilon \simeq 12.3$, $B_4=0$. When $ \epsilon \simeq 12.3$,  $B_4$ decreases sharply to $\simeq -7 $, indicating that the distribution of the densities deviates from Gaussian, revealing MIPS. The red points correspond to the activities of the systems depicted in the snapshots and the local density distributions.}}  
    \label{plot::BULK_L400_phi=0,1__density_frequency}
\end{figure}
%figure---------------------------------------------- 

%figure---------------------------------------------- PHASE DIAGRAM 
\begin{figure}[t!]
        %\captionsetup{justification=raggedright,singlelinecheck=false}
        \includegraphics[width=0.9\columnwidth]{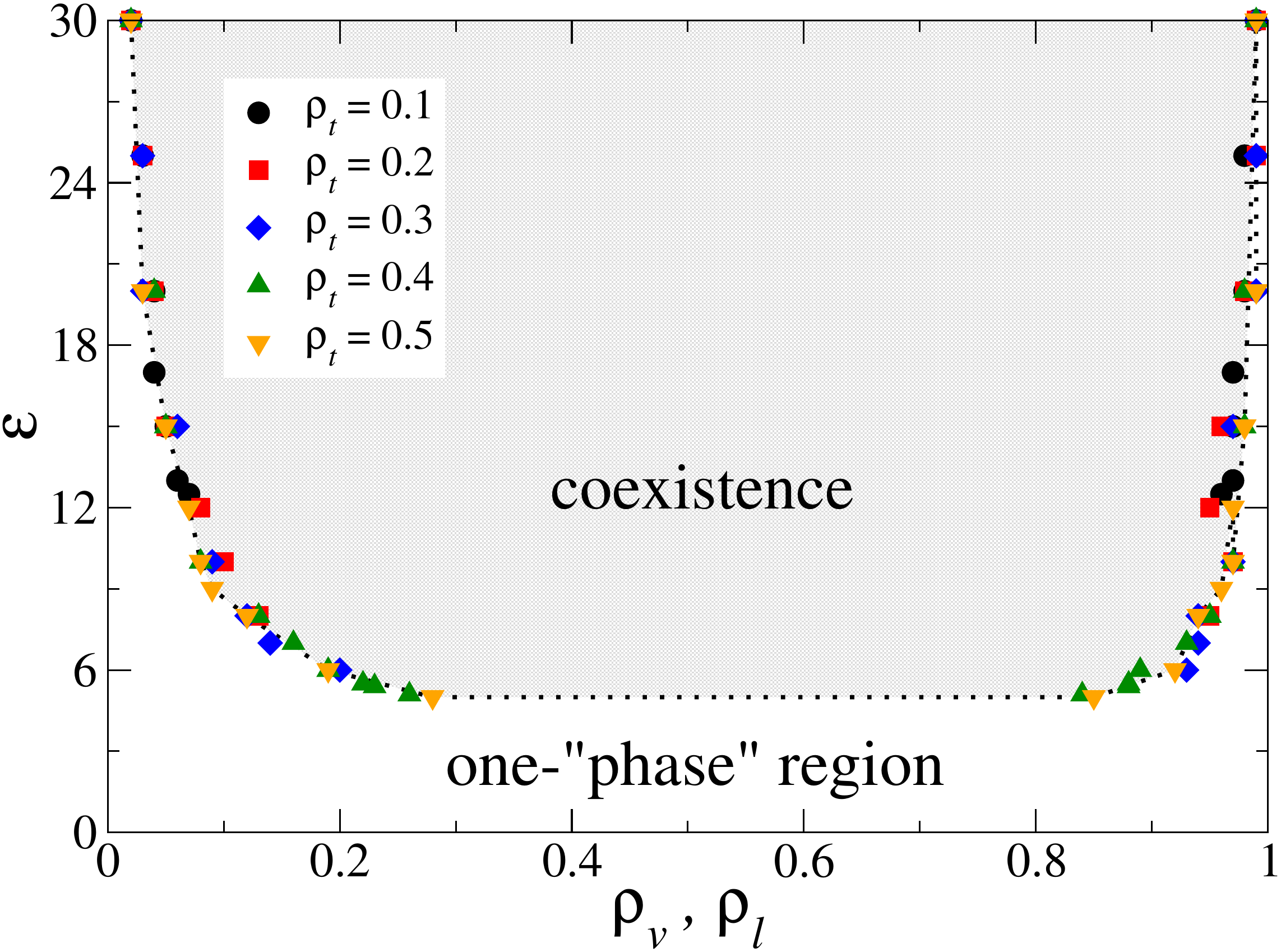}
        \caption{\small{Activity $\epsilon$ vs density, $\rho_l$ and $\rho_v$, phase diagram for $L=200$ and $D_r=0.1$ and different number of particles $N$ or total density $\rho_t$. The black dotted line depicts the liquid and vapour co-existing densities given by the maxima of the local density distributions for a system with total density $\rho_t=0.5$. The points correspond to the maxima of the density distributions for systems with different total densities. As expected, the coexisting densities do not depend on the total density. }}
    \label{plot::Phase_Diagram}
\end{figure}
%figure---------------------------------------------- \

We start by reporting the results for the simulations of the phase diagram of the model on a lattice with $L_x=L_y=L=200$. The simulation is run up to a KMC time $t= 2^{18}$, which is sufficient to reach the stationary state. The rotational diffusion coefficient was set to $D_r=0.1$ and kept fixed. The number of particles or total density is fixed and the activity $\epsilon$ is increased from the passive value 1 up to 30. We repeated the simulation for a range of total densities (number of particles). The bulk phase diagram of the model is a prerequisite for the investigation of the wetting behaviour reported in section \ref{sec.wetting}.

The top panel of Fig. \ref{plot::BULK_L400_phi=0,1__density_frequency} illustrates snapshots of a system with total density $\rho_t=0.1$. By increasing the activity a transition from a single phase (vapour) to a two-phase region (vapour and liquid) is observed. In order to calculate the coexisting densities we partition the lattice into sub-lattices of size $10 \times 10$, and plot in Fig.~\ref{plot::BULK_L400_phi=0,1__density_frequency}(a) the local density distribution at different activities. %This also allows the distinction between the single and the two-phase regions. 
The distribution exhibits one peak at $\rho=\rho_t$, in the single phase region, and two peaks, one at low and the other at high density, in the two-phase region. In Fig. \ref{plot::BULK_L400_phi=0,1__density_frequency}(a), only one peak is observed at  $\epsilon=2$, while above $\epsilon \geq 12.3$ a second peak is seen in the high density region, in line with the clusters depicted in the snapshots of the top panel. This threshold $\epsilon$ corresponds to the activity where the isochore $\rho_t=0.1$ hits the two-phase boundary.      

In the single phase region the local density distribution is (nearly) Gaussian and it becomes double peaked in the two-phase region. The fourth-order Binder cumulant ($B_4$) \cite{Binder1981} gives a quantitative measure of this change through the kurtosis. In the single-phase region, the kurtosis is nearly zero as a single Gaussian peak at $\rho_t = 0.1$ is observed. As the activity increases, the local density distribution changes and $B_4$ decreases signalling the appearance of a second peak in the high density region. $B_4$ is given by:

%Equatin---------------Binder
\begin{equation}
    B_4= 1- \frac{1}{3} \frac{\langle (\rho - \rho_t)^4 \rangle }{\langle(\rho - \rho_t)^2 \rangle ^2} ,    
\end{equation}
%Equation------------------------------
where $<.>$ is an average over the sub-lattices. The calculation of $B_4$ (for the final configuration) at each activity gives the threshold activity, that is the activity where the isochore hits the two-phase boundary. As shown in Fig. \ref{plot::BULK_L400_phi=0,1__density_frequency} $b)$, for activities less than $\epsilon=12.3$ the system is in a single phase with $B_4=0$. At an activity close to $12.3$ the Binder cumulant changes from zero to $\simeq -7$, indicating that the isochore $\rho_t=0.1$ hits the MIPS phase boundary, and the system phase separates, in line with both the snapshots and the local density distributions depicted in Fig. \ref{plot::BULK_L400_phi=0,1__density_frequency}. 

We have run simulations for total densities $\rho_t=\left(0.1, 0.2, 0.3, 0.4, 0.5\right)$ and plot the full phase diagram in Fig.~\ref{plot::Phase_Diagram}. As expected, the densities of the coexisting vapour ($\rho_{v}$) and liquid phases ($\rho_{l}$) depend only on the activity. Below a critical activity $\epsilon \approx 6$ the system does not phase separate, in line with previous results \cite{Partridge2018, Klamser2018}.

\subsection{Wetting of a solid surface}\label{sec.wetting}

\begin{figure*}
                  \includegraphics[width=0.9\textwidth]{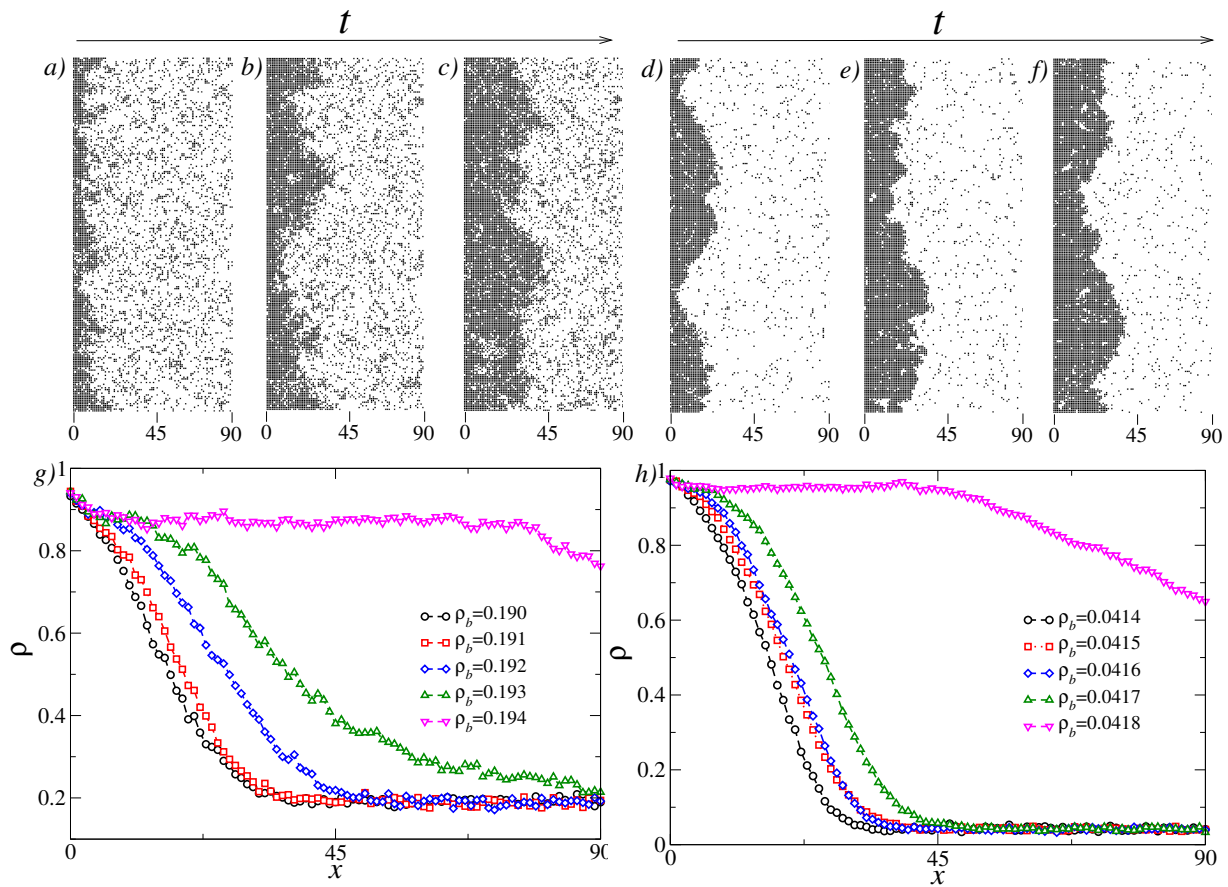}
       \caption{Snapshots of the time evolution of wetting films of active particles at a solid surface (located on the left of the snapshots) for a system with $L_x = 1000$. In { \it a)-c) } $\epsilon = 6, \rho_b = 0.193$, and in {\it d)-f)} $\epsilon = 18, \rho_b = 0.0417$. Laterally averaged number density as a function of the distance $x$ to the surface at {\it g)} $\epsilon = 6$, and {\it h)} $\epsilon = 18$; different curves correspond to different values of the density in the bulk region $V_b$ with size $L_b = 200$.  In addition to the steady state density profiles, unsteady ones are also shown: magenta down triangles in {\it g)} and {\it h)}. For these unsteady systems the liquid-vapour interface moves with a constant velocity, see Figs.~\ref{plot::h_vs_time}{\it a)} and {\it b)} below. Each profile is the result of averaging over 10 different realizations and over the two surfaces. The solid green curves correspond to the fitted inverse tangent model profile, and the vertical dashed line marks the locations of their inflection points $d_{inf}$. The location of the inflection points on the fitted curves provides our operational definition of the thickness of the wetting film. }
   \label{plot::densitry_profiles}
\end{figure*}
%figure----------------------------------------------

%figure---------------------------------------------- 

In wetting of equilibrium one-component systems in contact with a planar confining surface, one usually monitors the behaviour of the surface excess
 coverage $\Gamma$, or the thickness $h$ of the adsorbed liquid films, upon approaching the liquid-vapour coexistence line $\mu_{coex}(T)$ 
 from the vapour side while keeping the temperature $T$ constant \cite{dietrich1988}; here $\mu_{coex}(T)$ is the $T-$dependent value of chemical
  potential  $\mu$ at coexistence. In practical terms this amounts to calculating $\Gamma$ or $h$ as functions of the vapour undersaturation  $\Delta \mu \equiv \mu_{coex}(T) - \mu > 0$ at fixed $T$ \cite{dietrich1988,tasinkevych2006,Tasinkevych2007}. For $T$ sufficiently lower than the wetting temperature $T_w$, $h$ remains microscopically thin when $\Delta \mu \rightarrow 0^+$, however, $h \rightarrow \infty$ in this limit for $T \ge  T_w$, realising the  so-called complete wetting scenario. The specific form of this divergence depends on the range of the solid-fluid and fluid-fluid intermolecular interactions, and when both are short ranged $h\sim-\log(\Delta\mu)$ \cite{dietrich1988}.

%figure---------------------------------------------- FILM GROWTH
\begin{figure*}
       %\captionsetup{justification=raggedright,singlelinecheck=false}
       \vspace{-0.cm}
       \includegraphics[width=0.9\textwidth]{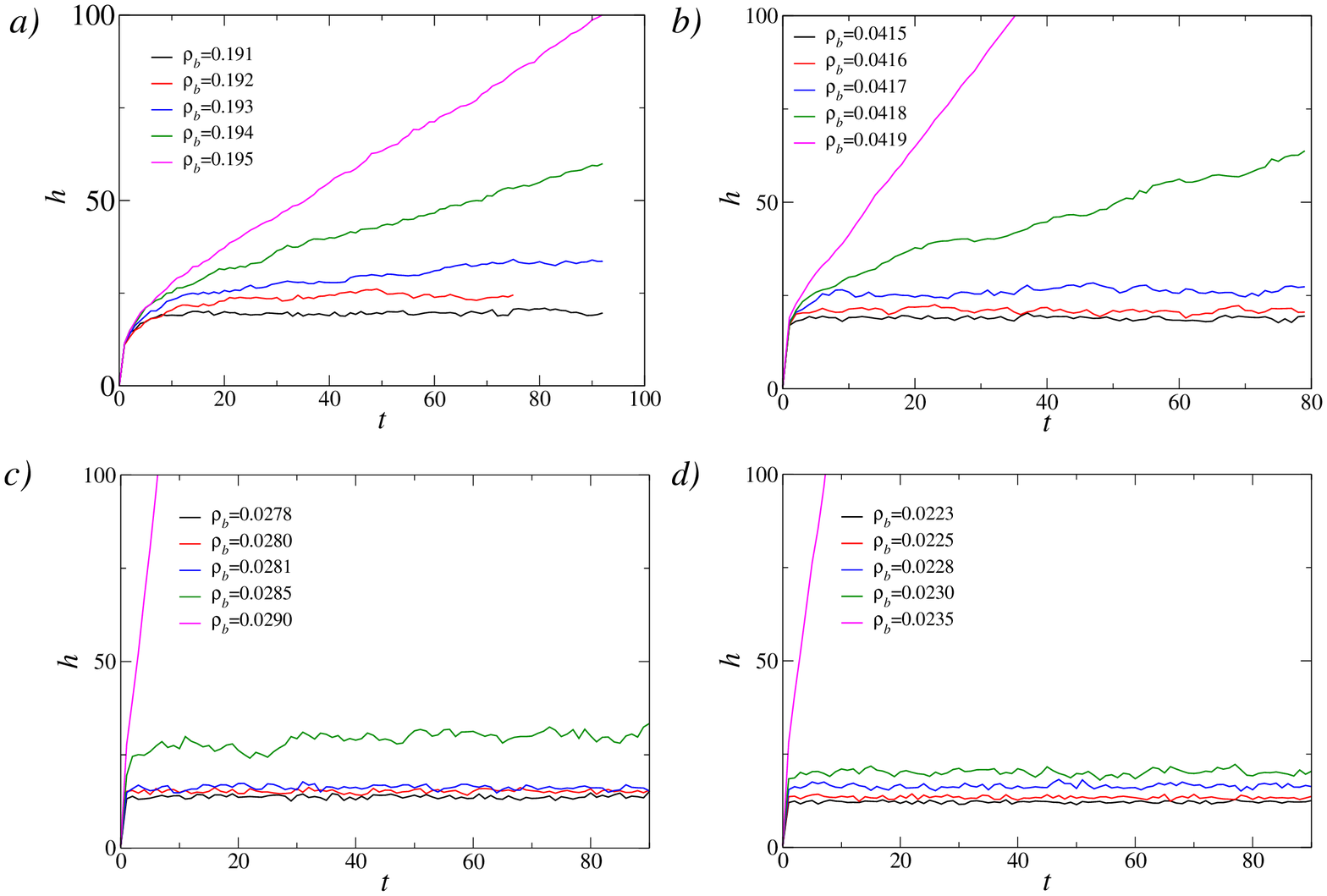}
       \vspace{-0.0cm}
       \caption{\small{ Thickness $h$ of the wetting liquid film, averaged over $L_y$,  as a function of time for several values of the bulk density $\rho_b$: \textit{{a)}} $\epsilon=6$; \textit{{b)}} $\epsilon=18$; \textit{{c)}} $\epsilon=25$; \textit{{d)}} $\epsilon=30$. In all cases $L_x = 1000$ .}}
   \label{plot::h_vs_time}
\end{figure*}
%figure----------------------------------------------

For all the results reported in the previous section, we carried out simulations at fixed volume and number of particles, i.e. an analog of the canonical ensemble of equilibrium statistical mechanics. In this section we implement an ensemble with fixed volume, but varying number of particles, which is supposed to mimic the grand canonical ensemble of equilibrium systems. To this end in the middle of the simulation box we consider a ``bulk'' region $V_b = L_b \times L_y$, where $L_b$ denotes the extension of the region in the direction normal to the confining surfaces. In this region the average number density of active particles is kept at a constant value $\rho_b < \rho_{v}$ in the course of the simulations, where $\rho_{v}$ is the density of the vapour phase at coexistence (see Fig.~\ref{plot::Phase_Diagram}). The constant density criterion is achieved by periodic insertion/removal of the  required number of randomly chosen particles into/from $V_b$. The other two regions which are adjacent to the confining surfaces, are free from such a constraint and are allowed to have a varying number of particles, as it is in these regions that the growth of the wetting films will occur. The wetting films grow through the supply of incoming particles from $V_b$, and in the steady state, particles are no longer inserted into $V_b$ and the film reaches a finite thickness. The extension of $V_b$ in the $x-$direction was chosen large enough to minimize the interactions between the liquid-vapour interfaces that form at the surfaces, as well as the interaction of these interfaces with the boundaries of the bulk region $V_b$.  Finally, in the $y-$direction we apply periodic boundary conditions for all the three slabs of the simulation box.
 
This method resembles the approach often used in density functional modeling of wetting \cite{singh2015,giacomello2019}. Here, one assumes that sufficiently far away from the wetting surface the fluid is homogeneous with the density given by the bulk density in the reservoir, characterised by the chemical potential $\mu_{coex}$ that is constant. Consequently, in the numerical calculations the fluid density at the wetting surface is fixed at the value of the bulk liquid at the same chemical potential $\mu_{coex}$  \cite{singh2015,giacomello2019}.

The objective of this section is to determine the steady state thickness $h$ of the liquid wetting films as a function of $\rho_b$ at different values of the particle activity $\epsilon$. The final aim is to clarify whether there exists a threshold of the activity which delimits a regime where the thickness $h$ of the wetting films remains finite at all values of $\rho_b \le \rho_{v}$ from a regime where $h$ diverges as $\rho_b \rightarrow \rho_{v}$.

%figure---------------------------------------------- FILM GROWTH
\begin{figure*}
       %\captionsetup{justification=raggedright,singlelinecheck=false}
       \vspace{-0.cm}
       \includegraphics[width=0.9\textwidth]{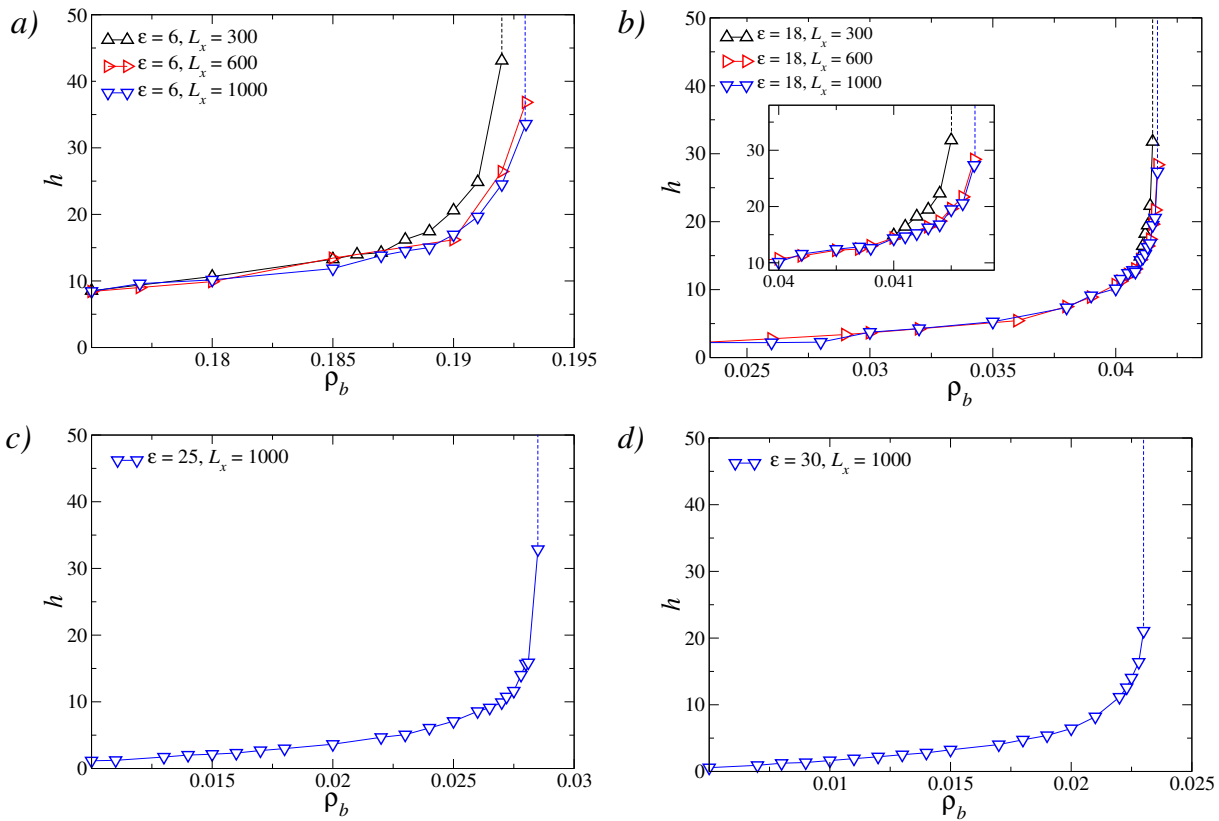}
       \vspace{-0.0cm}
       \caption{Thickness $h$ of the wetting film, averaged over $L_y$,  as a function of $ \rho_b$: \textit{a)} $\epsilon=6$; \textit{b)} $\epsilon=18$; \textit{c)} $\epsilon=25$; \textit{d)} $\epsilon=30$. For $\epsilon = 6, 18$, three different system sizes $L_x = 300, 600, 1000$ are shown. The size of the bulk region is $L_b = 100$ for $L_x=300$, and $L_b = 200$ for $L_x=600$, and 1000. }
   \label{plot::wetting_avg_thick}
\end{figure*}
%figure----------------------------------------------

In Figs.~\ref{plot::densitry_profiles}(a) and (b) we show typical snapshots of the system evolution for two values of the particle activity $\epsilon = 6$ and $\epsilon = 18$, which are close and far from the MIPS critical point, respectively; $\rho_b$ is slightly smaller than $\rho_{v}$. The structure of the wetting liquid film exhibits large density fluctuations, note the presence of ``'bubble''-like domains, while the liquid-vapour interface is extremely rough with several interface overhangs. The laterally averaged density profiles $\rho(x)$ as functions of the distance $x$ to the surface are reported in Fig.~\ref{plot::densitry_profiles}(g) and (h) for several values of $\rho_b$ and for two particle activities. In both cases,  the ``liquid shoulder'' of $\rho(x)$ grows with $\rho_b$ reaching the value of $x \approx 30$ for the largest $\rho_b$, reported, and decays towards $\rho=\rho_b$ at larger values of $x$.  

The time evolution of $h$ is presented in Fig.~\ref{plot::h_vs_time} for different $\epsilon$ and $\rho_b$. For $\rho_b < \rho_{v}$, the film thickness reaches the steady state relatively fast for large values of $\epsilon$ regardless of the value of $\rho_b<\rho_{v}$, see black, red, and blue lines in 
Fig.~\ref{plot::h_vs_time}(b), (c), and (d) as well as the green curves in Fig.~\ref{plot::h_vs_time}(c) and (d). By contrast, for activities close to the MIPS critical point, $\epsilon=6$, the time needed to reach the steady state grows considerably as $\rho_b$ approaches $\rho_{v}$, see the blue line in Fig.~\ref{plot::h_vs_time}(a). We also show several examples of unsteady 
film growth where $h$ increases with a constant speed, magenta curves in  Fig.~\ref{plot::h_vs_time} and green curves in Fig.~\ref{plot::h_vs_time}(a) and (b). In these cases, $\rho_b$ is supposedly larger than $\rho_{v}$, active particles start to form extensive and dense liquid films which grow until reaching the boundary of the bulk region $V_b$, and stopping afterwards.

After analysing the time evolution of the film thickness, we calculate the steady state $h$ as a function of the particle density $\rho_b$ in the bulk region $V_b$. The results are shown in Fig.~\ref{plot::wetting_avg_thick} for several activities and system sizes $L_x$. The vertical dashed lines mark the boundary $\rho^{*}_{b}$ between the steady states (on the left of the line) and unsteady film growth with constant velocity (on the right of the line). The value of  $\rho^{*}_{b}$ decreases with decreasing $L_x$, due to the proximity of the liquid-vapour interface and the boundary of $V_b$. Indeed, at  $\epsilon = 6$, the average length of the straight portions of the particle trajectories is $\epsilon/D_r = 60$, while the size of the bulk region is $L_b = 100$ for $L_x = 300$. Under these circumstances, when $h\sim 40$, see black up-triangles in  Fig.~\ref{plot::wetting_avg_thick} (a), the liquid in the film starts to ``feel'' the particles in the bulk region, which can enter directly into the liquid film, due to the large persistence length of their trajectories. We expect that $\rho^{*}_{b} \rightarrow \rho_{v}$ as $L_x \rightarrow \infty$.
Surprisingly, $h(\rho^{*}_b)\approx 40$, which is far from being microscopic, irrespective of the value of $\epsilon$ near, $\epsilon = 6$, or far, $\epsilon = 30$, from the MIPS critical point. Additionally, inspection of the profiles $h(\rho_b)$ reveals that their functional form may be fitted by a double logarithmic divergence $ - \log \left [ \log\left (\rho_{v} - \rho_{b}\right) \right ]$, as equilibrium wetting with short range attraction suggests, for all the values of $\epsilon$ investigated (note that close to the bulk coexistence the undersaturation is $\Delta \mu \propto \Delta \rho \equiv \rho_{v}-\rho_b$). The precise form of the divergence of $h(\rho_b)$ is not relevant, but the simulations suggest that $h(\rho_b)$ {\bf does} diverge in the whole range of $\epsilon$. This means that our model predicts complete wetting of a solid surface at all activities.

%--------------------------- CONCLUSIONS-------------------------------- 

\section{Discussion}

Active particles belong to a class of non-equilibrium systems with a persistent local entropy production, which violates detailed balance. This gives rise to a novel type of collective behaviour such as the coexistence of vapour- and liquid-like steady states for active particles with repulsive interactions only.  Similarly to bacteria, active particles tend to accumulate at confining surfaces forming dense adsorbed films. This naturally leads to the question of whether active particles exhibit an analogue of the wetting transitions known for fluids at equilibrium. 

For ABPs at a planar surface, Wittmann and Brader adopted an effective equilibrium approach \cite{Farage2015} and constructed an effective activity-dependent free energy functional with effective activity-induced particle-particle and particle-surface interactions \cite{Wittmann2016}. As the particle activity increases, the effective interactions develop deeper and deeper minima that shift towards smaller separations. For repulsive particles at a repulsive surface, this effective density functional theory predicts the divergence of the wetting film thickness, upon approaching the liquid-vapour coexistence (from the vapour side) at activity close to the MIPS critical point. However, higher values of the activity (further away from the critical point) were not investigated leaving the existence of a wetting singularity at coexistence an open question. Moreover, it is known that the mapping to an effective equilibrium system fails beyond the linear response regime and misses essential non-equilibrium properties of the active particles \cite{Rein2016}.

A similar conclusion that  ``equilibrium'' is not the whole story was made in \cite{Perez-Gonzalez2019}, where the wetting behaviour of epithelial tissues on different substrates was investigated both experimentally and theoretically. The main result of this study is that the wetting properties of the active tissue are determined by the competition of forces across the entire tissue monolayer adjacent to the substrate. This is in sharp contrast to the case of passive fluids at equilibrium, where spreading of
a sessile drop is determined by the forces ``operating'' at the contact line \cite{Perez-Gonzalez2019, Gennes1985}.

Here we have considered a lattice active gas model with short ranged repulsion in order to study active wetting.   
The dynamics of the model has been resolved using the Kinetic Monte Carlo approach. First, we have determined the bulk coexistence curve between the dense liquid-like and dilute vapour-like steady states. For the bulk simulations we have used systems with periodic boundary conditions and fixed number of particles, an analog of the canonical ensemble of equilibrium systems. Next, we implemented an ensemble with a varying number of particles. This quasi grand canonical ensemble, was used  to investigate the wetting of a solid-vapour interface by the liquid ``phase'', i.e. the formation and growth of liquid wetting films between a planar solid surface and the vapour phase. Extensive simulations demonstrated, for all the values of the activity considered, a signature of diverging wetting films as the system is brought towards the coexistence curve from the vapour side. In other words we have not found a wetting singular point at the coexistence curve, which separates an incomplete wetting regime from a complete wetting one.  Instead, our results suggest that the complete wetting scenario persists in the whole range of activities considered. 

In the present study we have not investigated the structure of the wetting films. However, very recent simulations of active lattice gas models in the bulk, \cite{Shi2020} demonstrated that the liquid phase at MIPS is permeated by vapour bubbles of different sizes distributed algebraically. Moreover, at large enough system sizes or global densities, the vapour phase may be totally consumed by these bubbles forming, as the authors of  refs.~\cite{Tjhung2018,Shi2020} call it, ``microphase-separated bubbly liquids''.  One possible extension of this study, is to investigate whether a similar bubbly structure is found in the wetting films, and if yes, how this would affect, if at all, the complete wetting scenario advocated here. Another direction of study is a detailed investigation of the dynamical behaviour of the liquid-vapour interface at the growing wetting film, and in particular its roughness.

\section{Acknowledgments}

We acknowledge financial support from the Portuguese Foundation for Science and Technology (FCT) under Contracts no. PTDC/FIS-MAC/28146/2017 (LISBOA-01-0145-FEDER-028146), CEECIND/00586/2017, UIDB/00618/2020, UIDP/00618/2020, and IF/00322/2015.

\bibliographystyle{unsrt}
\bibliography{Active_matter}

\end{document}